\documentclass[10pt,twocolumn,conference]{IEEEtran}

\IEEEoverridecommandlockouts

\usepackage{graphicx}
\usepackage{amsmath}
\usepackage{amssymb}
\usepackage[caption=false]{subfig}
\usepackage[noadjust]{cite}
\usepackage{float}
\usepackage{algorithm}
\usepackage{bibentry}
\usepackage{balance}
\usepackage{algorithm}
\usepackage{algorithmicx}
\usepackage{algpseudocode}
\usepackage{xcolor}
\usepackage{subfig}

\graphicspath{{img/}}

\begin{document}

\bstctlcite{IEEEexample:BSTcontrol}

\title{Precoder Design for mmWave UAV Communications with Physical Layer Security\thanks{This work is supported in part by the INL Laboratory Directed Research Development (LDRD) Program under DOE Idaho Operations Office Contract DEAC07-05ID14517.}}

\author{
\IEEEauthorblockN{Sung Joon Maeng$^*$, Yavuz Yap{\i}c{\i}$^*$, \.{I}smail G\"{u}ven\c{c}$^*$, Huaiyu Dai$^*$, and Arupjyoti Bhuyan$^\dagger$}\IEEEauthorblockA{$^*$Department of Electrical and Computer Engineering, North Carolina State University, Raleigh, NC\\
$^\dagger$Idaho National Laboratory, Idaho Falls, ID\\
\{smaeng, yyapici, iguvenc, hdai\}@ncsu.edu, arupjyoti.bhuyan@inl.gov}}

\maketitle

\begin{abstract}
The integration of unmanned aerial vehicles (UAVs) into the terrestrial cellular networks is envisioned as one key technology for next-generation wireless communications. In this work, we consider the physical layer security of the communications links in the millimeter-wave (mmWave) spectrum which are maintained by UAVs functioning as base stations (BS). In particular, we propose a new precoding strategy which incorporates the channel state information (CSI) of the eavesdropper (Eve) compromising link security. We show that our proposed precoder strategy eliminates any need for artificial noise (AN) transmission in underloaded scenarios (fewer users than number of antennas). In addition, we demonstrate that our nonlinear precoding scheme provides promising secrecy-rate performance even for overloaded scenarios at the expense of transmitting low-power AN.     
\end{abstract}


\section{Introduction}
Recently, a growing demand of unmanned aerial vehicles (UAVs), also known as drones, in plentiful applications gives rise to the necessity for studying UAV wireless communications. Due to the flexibility in the size and the mobility, UAVs are used in various applications in civilian, for instance, search-and-rescue operation during the emergency and droned based delivery service, as well as in military such as airspace surveillance \cite{hayat2016survey}. Moreover, such characteristics of UAV enables UAV to be deployed as a flying base station (UAV-BS) \cite{rupasinghe2018non}, and UAV-BS operates to improve coverage and connectivity with ground users.

UAV wireless cellular networks is vulnerable to illegitimate malicious receivers due to the broadcast nature. Traditionally, cryptographic encryption is implemented at the network and application layer to ensure security \cite{mukherjee2014principles}. This technique is based on the limited computational complexity of adversaries. The physical layer security (PLS) has recently received attention as an effective complement. The wire-tap channel is considered in \cite{wyner1975wire}, which consists of a transmitter, intended receivers, and an eavesdropper.

UAV trajectory and transmit power are optimized to maximize achievable secrecy rate on the wire-tap channel in \cite{cui2018robust}. A friendly jammer UAV, which transmit artificial noise (AN) for the purpose of secure communications is introduced in \cite{li2019cooperative}, and the trajectory and the transmit power of both source and jammer UAV are optimized. In \cite{goel2008guaranteeing}, the BS equipped with multiple antenna transmits artificial noise as a source combined with data. In \cite{zhu2015linear}, precoders for both data and AN are designed by well-known linear precoding with legitimate user channels knowledge for the secure transmission.

In this paper, we study the PLS of broadband coverage provided by a UAV-BS in mmWave frequency spectrum. In particular, we consider transmission of AN in a smart fashion, which takes into account not only the legitimate user channels but also limited or full feedback on the Eve CSI. We propose two precoding strategies that design precoders for both the data and AN signal in linear and nonlinear fashions, which utilize the available Eve CSI. We numerically show that the proposed nonlinear precoding scheme with full Eve CSI offers a promising secrecy-rate performance for both underloaded (few users) and overloaded (many users) scenarios, where the need for AN is even completely eliminated for underloaded scenarios. The proposed linear precoding scheme also outperforms the conventional strategies not using Eve CSI, and is very effective especially for underloaded scenarios even with limited Eve CSI.    

\textit{Notation:} $\|\cdot\|^2$  and $\|\cdot\|_\mathsf{F}$ denote the Euclidean and Frobenious norm operators, respectively.  $\textbf{I}$ is the identity matrix, $\mathbb{C}$ is complex numbers. $\mathcal{CN}(0, \alpha\textbf{I})$ stands for complex Gaussian distribution with zero mean and covariance matrix $\alpha\textbf{I}$. $\operatorname{Im} \{\textbf{X}\} $ represents the imaginary parts of matrix $\textbf{X}$. $\mathcal{U}[x_1,x_2]$ denotes uniformly random distribution with the range from $x_1$ to $x_2$. $\text{null}(\textbf{X})$ represents null space of matrix $\textbf{X}$.

\section{System Model} \label{sec:system}

We consider a mmWave communications scenario where a single UAV-BS flying at an altitude of $h_\mathsf{uav}$ serves $K$ single-antenna users with the index set $\mathcal{K} \,{=}\, \{1,\dots,K\}$. The users are assumed to be deployed uniformly on the ground with $d_k$ being the horizontal distance (on the ground) between the UAV-BS and $k$-th user such that $d_k \,{\sim}\,\mathcal{U}[d_\mathsf{min},d_\mathsf{max}]$. The corresponding communication is being wiretapped by a single passive eavesdropper (Eve). The UAV-BS and Eve are equipped with vertically positioned uniform linear array (ULA) having $N$ and $M$ antenna elements, respectively. We assume that the UAV-BS serves the users within the same spectral (i.e., time and frequency) resources (without any scheduling) through a \textit{multiuser all-digital precoding} scheme. In order to improve the resiliency against possible Eve operations and enable secure transmission, the UAV-BS transmits AN along with the user messages using the same spectral resources.

The received signal at the $u$-th user is given by 
\begin{align}\label{eq:user_y}
    y_u&=\sum_{k=1}^{K}\sqrt{\mathsf{P}_\mathsf{Tx}}\textbf{h}_u^{\rm H}\textbf{w}_ks_k+\sum_{i=1}^{Z}\sqrt{\mathsf{P}_\mathsf{Tx}}\textbf{h}_u^{\rm H}\textbf{v}_iz_i+n_u,
\end{align}
where $\mathsf{P}_\mathsf{Tx}$ is the transmit power at the UAV-BS, $\textbf{h}_u\in\mathbb{C}^{N\times 1}$ is the $u$-th user's channel vector, $\textbf{W}=[\textbf{w}_1,\dots,\textbf{w}_K]\in\mathbb{C}^{N\times K}$ is the aggregate data precoder matrix for user messages with $\textbf{w}_k$ being the precoder for the $k$-th user message $s_k$, $\textbf{V}=[\textbf{v}_1,\dots,\textbf{v}_K]\in\mathbb{C}^{N\times Z}$ is the precoder matrix for the AN transmission with $\textbf{v}_i$ being the precoder associated with the $i$-th AN vector $z_i$, $Z \,{\leq}\, N \,{-}\, K$ represents the degrees of freedom available for the AN transmission, and $n_u$ is the respective observation noise following $\mathcal{CN}(0, \sigma_n^2)$. Similarly, the received (wiretapped) signal at the Eve is given as
\begin{align}\label{eq:user_eve}
    \textbf{y}_\mathsf{e}&=\sum_{k=1}^{K}\sqrt{\mathsf{P}_\mathsf{Tx}}\textbf{H}_\mathsf{e}^{\rm H}\textbf{w}_ks_k+\sum_{i=1}^{Z}\sqrt{\mathsf{P}_\mathsf{Tx}}\textbf{H}_\mathsf{e}^{\rm H}\textbf{v}_iz_i+\textbf{n}_\mathsf{e},
\end{align}
where $\textbf{H}_\mathsf{e}\in\mathbb{C}^{N\times M}$ is the channel matrix of Eve, and $\textbf{n}_\mathsf{e}\in\mathbb{C}^{M\times 1}$ is the observation noise following $\mathcal{CN}(0, \sigma_n^2\textbf{I})$.

\begin{figure}[!t]
	\centering
	\vspace{-0.0in}
	\includegraphics[width=0.47\textwidth]{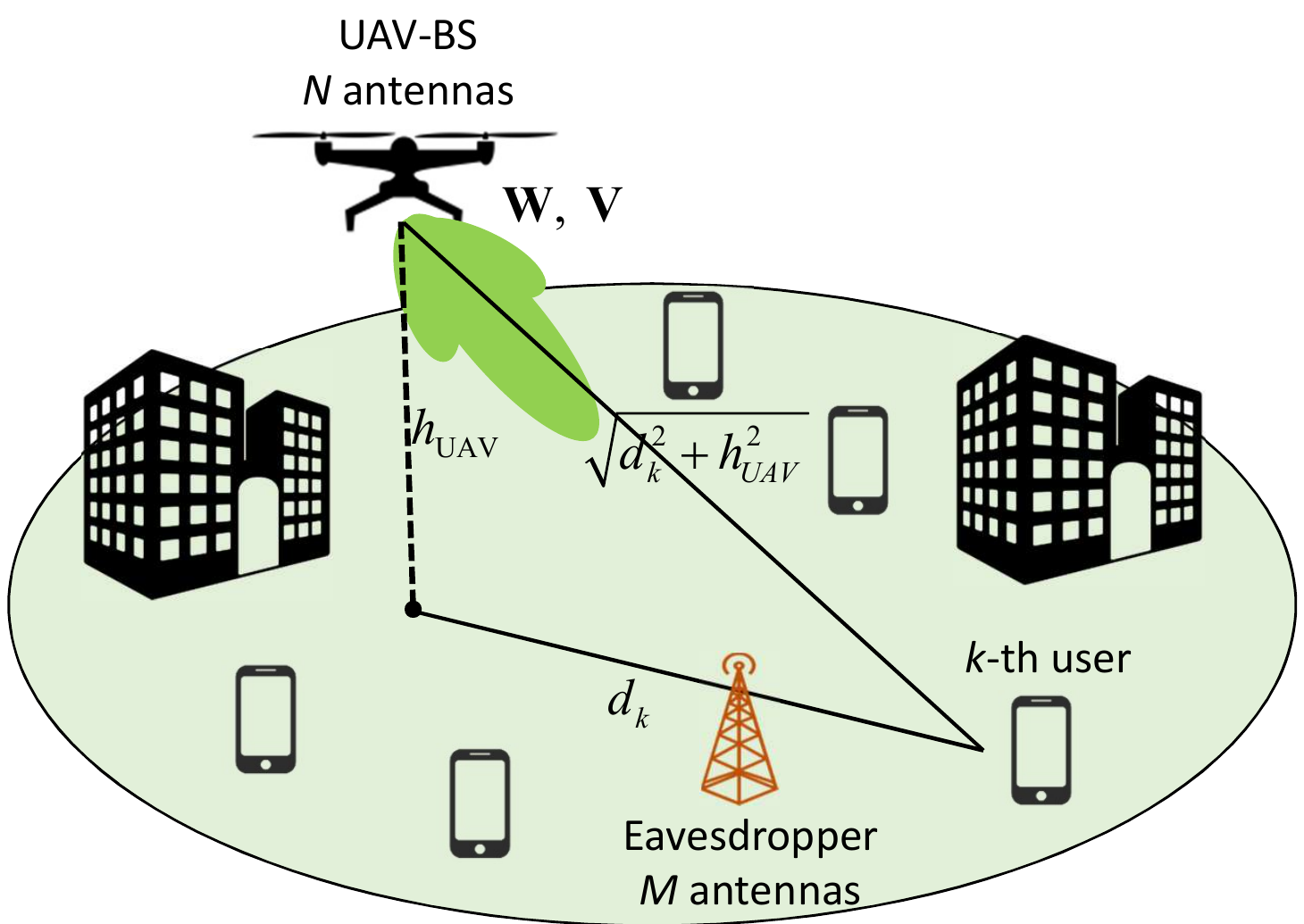}
	\vspace{-0.0in}
	\caption{Illustration of system scenario with UAV-BS wireless networks.}
	\label{fig:illu}
\end{figure}

The channel between the UAV-BS and $k$-th user is given as
\begin{align}\label{eq:channel_user}
    \textbf{h}_k = \sqrt{ \frac{ N }{ L_\mathsf{p} } } \sum^{L_\mathsf{p}}_{\ell=1} \frac{ \alpha_{k,\ell} }{ \mathsf{PL} \left( d_k \right) } \textbf{a}_N(\theta_{k,\ell}),
\end{align}
where $L_\mathsf{p}$ is the number of multipaths, $\alpha_{k,\ell}$ is the small-scale fading of the $k$-th user along the $\ell$-th path being standard complex Gaussian, $\mathsf{PL} \left( d_k \right)$ is the path loss for the $k$-th user, and $\theta_{k,\ell}$ is the angle of departure (AoD) for the $\ell$-th path with the array steering vector given as  
\begin{align}\label{eq:steering_vector}
    \textbf{a}_N(\theta)&=\frac{1}{\sqrt{N}}\left[1\; e^{-j\frac{2\pi d_{\rm s}}{\lambda}\sin\theta}\;\dots\; e^{-j\frac{2\pi d_{\rm s}}{\lambda}(N-1)\sin\theta}\right]^{\rm T},
\end{align}
where $d_{\rm s}$ and $\lambda$ are the antenna spacing and wavelength of the carrier frequency $f_\mathsf{c}$. Note also that we assume that the AoD is Laplacian distributed around the line-of-sight (LoS) angle $\theta_\mathsf{LoS}$ with the angle spread of $\Delta$. In a similar fashion, the channel between the UAV-BS and Eve is given as
\begin{align}\label{eq:channel_eve}
    \textbf{H}_\mathsf{e} = \sqrt{\frac{NM}{L_\mathsf{p}}} \sum^{L_\mathsf{p}}_{\ell=1} \frac{ \alpha_{\mathsf{e},\ell} }{ \mathsf{PL} \left( d_\mathsf{e} \right) } \textbf{a}_N(\theta_{\mathsf{e},\ell})\textbf{a}_M^{\rm H}(\psi_{\mathsf{e},\ell}),
\end{align}
where $\alpha_{\mathsf{e},\ell}$ is the small-scale fading along the $\ell$-th path, $d_\mathsf{e}$ is the horizontal distance between the UAV-BS and Eve following $\mathcal{U}[d_\mathsf{min},d_\mathsf{max}]$, and $\theta_{\mathsf{e},\ell}$ and $\psi_{\mathsf{e},\ell}$ are the AoD and angle of arrival (AoA) associated with the $\ell$-th path where AoA has the same distribution as AoD. 

\section{Precoder Design for Secure Transmission}

In this section, we present the achievable secrecy rate, and overview conventional precoder design schemes associated with the data and AN transmission to enable secure communications. To this end, we first express the signal-to-interference-plus-noise ratio (SINR) of the $u$-th user as follows 
\begin{align}\label{eq:SINR_u}
    \!\!\mathsf{SINR}_u = \frac{\textbf{w}_u^{\rm H}\textbf{h}_u\textbf{h}_u^{\rm H}\textbf{w}_u}{\sum_{k\neq u}^{K} \textbf{w}_k^{\rm H}\textbf{h}_u\textbf{h}_u^{\rm H}\textbf{w}_k + \sum_{i=1}^{Z} \textbf{v}_i^{\rm H}\textbf{h}_u\textbf{h}_u^{\rm H}\textbf{v}_i + \rho^{{-}1}},
\end{align}
where $\rho \,{=}\, \frac{ \sigma_n^2}{ \sqrt{\mathsf{P}_\mathsf{Tx}}}$ is the signal-to-noise ratio (SNR). In addition, the SINR of Eve is given as follows \cite{zhu2015linear}
\begin{align}\label{eq:SINR_eve}
    \mathsf{SINR}_\mathsf{e} = \textbf{w}_u^{\rm H}\textbf{H}_e(\textbf{H}_e^{\rm H}\textbf{V}\textbf{V}^{\rm H}\textbf{H}_e+\rho^{{-}1} \textbf{I})^{-1}\textbf{H}_e^{\rm H}\textbf{w}_u,
\end{align}
where \eqref{eq:SINR_eve} is equivalent to the maximum SINR at the MMSE combiner \cite{gao1998theoretical}.
We assume the worst-case that the Eve is able to decode and cancel the signals of UAV-BS except the signal of interest. The achievable secrecy rate of the $u$-th user is then given as
\begin{align}
     \mathsf{R}^\mathsf{sec}_u &= \left[\log(1+\mathsf{SINR}_u)-\log(1+\mathsf{SINR}_\mathsf{e})\right]^+, \label{eq:scrate}\\
     &= \left[\log \left( \frac{ 1+\mathsf{SINR}_u }{ 1+\mathsf{SINR}_\mathsf{e}} \right) \right]^+, \label{eq:scrate_v2}
\end{align}
where $[x]^+ = \max(0,x)$.

Assuming unit-energy overall precoding mechanism (i.e, $\|\textbf{W} \,\textbf{V}\|^2 \leq 1$), the power allocated to data and AN precoders is controlled as follows 
\begin{align}\label{eq:power_constraint}
    \|\textbf{W} \|^2 &= \phi, \\ \|\textbf{V}\|^2 &= 1-\phi,
\end{align}
where $\phi$ is the power splitting factor. The well-known  zero-forcing (ZF), and  regularized zero-forcing (RZF) precoding schemes are defined for the data precoding part as follows
\begin{align}
    \text{ZF:}\quad\Tilde{\textbf{W}} &= \textbf{H}(\textbf{H}^{\rm H}\textbf{H})^{{-}1}, \label{eq:precoder_zf_unnormalized}\\
    \text{RZF:}\quad\Tilde{\textbf{W}} &=(\textbf{H}\textbf{H}^{\rm H}+\beta\textbf{I})^{-1}\textbf{H}
    \label{eq:precoder_rzf_unnormalized}
\end{align}
where $\textbf{H} \,{=}\, \left[ \textbf{h}_1 \dots \textbf{h}_K \right]$ is the aggregate channel matrix. The final data precoder for the $k$-th user is given accordingly under the uniform power allocation assumption as follows
\begin{align} \label{eq:precoder_user_normalized}
    \textbf{w}_k&=\sqrt{\frac{\phi}{K\,\mathsf{tr}(\Tilde{\textbf{w}}_k^{\rm H}\Tilde{\textbf{w}}_k)}}\Tilde{\textbf{w}}_k,
\end{align}
where $\Tilde{\textbf{w}}_k$ is the $k$-th column of   $\Tilde{\textbf{W}}$.

The well-known strategy for designing AN precoder is \textit{null-space} precoding, which is defined as
\begin{align} \label{eq:precoder_eve_unnormalized}
    \Tilde{\textbf{V}}&=\text{null}(\textbf{H}^{\rm H}),
\end{align}
and the final precoder vectors are obtained through uniform power allocation as follows
\begin{align} \label{eq:precoder_eve_normalized}
    \textbf{v}_i &= \sqrt{\frac{1-\phi}{Z\,\mathsf{tr}(\Tilde{\textbf{v}}^{\rm H}_i\Tilde{\textbf{v}}_i)}}\Tilde{\textbf{v}}_i,
\end{align}
where $\Tilde{\textbf{v}}_i$ is the $i$-th column of   $\Tilde{\textbf{V}}$.

We would like to remark that the conventional data precoders in \eqref{eq:precoder_zf_unnormalized}-\eqref{eq:precoder_user_normalized} take into account the users only, and try to mitigate the interference coming from the multiuser communications (i.e., all the users are being served using the same spectral resources). In addition, the AN precoder in \eqref{eq:precoder_eve_unnormalized}-\eqref{eq:precoder_eve_normalized} proceeds to suppress the ability of Eve to wiretap the users' messages by simply putting noise toward null space of the user aggregate channel matrix (i.e., towards directions where no user is present). The conventional strategy, however, does not utilize the Eve channel state information (CSI) (i.e., $\textbf{H}_\mathsf{e}$) in either the data or AN precoding parts. In the next section, we propose a new precoder design strategy where the CSI of Eve is incorporated into the design problem.

\section{Eve-CSI-Aware Precoder Design}

In this section, we propose a new precoder design strategy for the data and AN signal assuming full and partial CSI (on the Eve's channel) separately. We first consider a linear scheme, which benefits from existing ZF and RZF precoders but incorporates also Eve CSI. We then propose a novel nonlinear precoding strategy which offers a promising secrecy-rate performance. 

\subsection{Linear Precoding with Full vs. Limited CSI}

In the AN transmission, the main goal is to put as much power as possible towards Eve direction while not degrading the legitimate user channels much. The AN transmission is conventionally designed to happen towards the directions within the null space of the user channel matrix $\textbf{H}$ since the CSI on the Eve channel is not benefited. We now propose linear precoder design strategy for the data and AN signal which employs not only the user channel matrix but also the Eve's CSI.

To this end, we first consider the singular value decomposition (SVD) of the channel matrix of Eve as follows
\begin{align}\label{eq:eve_channel_svd}
    \textbf{H}_\mathsf{e} &= \sum_{\ell=1}^{R}\lambda_\ell \textbf{u}_\ell \textbf{d}_\ell^{\rm H},
\end{align}
where $\lambda_\ell$ is the $\ell$-th singular value with $\lambda_1 \,{>}\, \dots \,{>}\,\lambda_R$, $\textbf{u}_\ell$ and $\textbf{d}_\ell$ are the left and right singular vectors, respectively, and $R$ is the rank of the channel matrix. Note that the mmWave channel is sparse in the angular domain so that most of the energy is focused around few directions. We therefore represent the channel by a rank-1 matrix as follows 
\begin{align} \label{eq:eve_channel_equivalent}
    \textbf{H}_\mathsf{e} \approx \lambda_1 \textbf{u}_1 \textbf{d}_1^{\rm H},
\end{align}
where the right singular vector can be interpreted as the dominant multipath direction towards Eve.

We consider a \textit{virtual} channel matrix which aggregates the user channel matrix and the dominant Eve direction described in \eqref{eq:eve_channel_equivalent}, which is given as follows 
\begin{align}\label{eq:agg_chnl}
    \textbf{G}&=\left[\textbf{H} \; \textbf{d}_1 \right].
\end{align}
Treating Eve as if it is a regular user, the respective ZF and RZF precoders are given as
\begin{align}
    \text{ZF:}\quad\textbf{F}&=\textbf{G}(\textbf{G}^{\rm H}\textbf{G})^{-1}\hspace{0.7in} \label{eq:proposed_zf}\\
    \text{RZF:}\quad\textbf{F}&=(\textbf{G}\textbf{G}^{\rm H}+\beta\textbf{I})^{-1}\textbf{G} \label{eq:proposed_rzf}
\end{align}
where $\beta$ is the regularization parameter, and $\textbf{F} \,{=}\, [\textbf{f}_1,\dots,\textbf{f}_{K{+}1}]$  is the aggregate precoding matrix. In this definition, $\textbf{f}_k$ is the precoder for the $k$-th user for $k \,{\in}\, \mathcal{K}$, and $\textbf{f}_{K{+}1}$ is the precoder for Eve. Note that in either ZF or RZF precoding scheme, the $k$-th user precoder $\textbf{f}_k$ allocates maximum power towards its desired user while suppressing the multiuser interference associated with not only the other users but also Eve. Similarly, $\textbf{f}_{K+1}$ operates to put more power towards dominant Eve direction \textit{while keeping the resulting interference at all the users minimum}. We therefore conclude that $\textbf{f}_{K+1}$ indeed represents an effective candidate for AN precoder. The resulting unnormalized data and AN precoders are given as follows
\begin{align}
    \Tilde{\textbf{W}} &= \left[\textbf{f}_1,\dots,\textbf{f}_K\right], \label{eq:proposed_data_precoder_unnormalized}\\ 
    \Tilde{\textbf{v}} &= \textbf{f}_{K+1}, \label{eq:proposed_eve_precoder_unnormalized}
\end{align}
which take the form after uniform power allocation as follows
\begin{align}
    \textbf{w}_k &= \sqrt{\frac{\phi}{K\mathsf{tr}(\Tilde{\textbf{w}}_k^{\rm H}\Tilde{\textbf{w}}_k)}}\Tilde{\textbf{w}}_k, 
    \label{eq:proposed_data_precoder_normalized} \\
    \textbf{v} & = \sqrt{\frac{1-\phi}{\mathsf{tr}(\Tilde{\textbf{v}}^{\rm H} \Tilde{\textbf{v}})}}\Tilde{\textbf{v}}. 
    \label{eq:proposed_eve_precoder_normalized}
\end{align}

When the UAV-BS does not have full CSI of Eve, but has a partial information in the form of Eve's location, the aggregate channel based on this relatively stationary LoS angle $\theta_\mathsf{LoS}$ is given as
\begin{align}\label{eq:agg_chnl_lim}
    \textbf{G}&=[\textbf{H}\;\textbf{a}_N(\theta_\mathsf{LoS})],
\end{align}
where $\textbf{a}_N(\theta_\mathsf{LoS})$ is the array steering vector described in \eqref{eq:steering_vector}. Although the limited feedback scheme in \eqref{eq:agg_chnl_lim} is not as accurate as the approximation in \eqref{eq:agg_chnl} relying on dominant Eve direction, the respective secrecy-rate performance is shown to be reasonable in Section~\ref{sec:results}.  

\subsection{Nonlinear Precoder Design with Full CSI}

\begin{table}[!t]
\renewcommand{\arraystretch}{1.1}
\caption{Simulation settings}
\label{table:settings}
\centering
\begin{tabular}{lc}
\hline
Parameter & Value \\
\hline\hline
Transmit power ($\mathsf{P}_\mathsf{Tx}$) & 30 dBm\\
Antenna element spacing ($d_{\rm s}$) & $\frac{\lambda}{2}$ \\
Number of antennas at UAV-BS ($N$) & $16$ \\
Number of antennas at Eve ($M$) & 4 \\
Number of users ($K$) & $\{4, 16, 32\}$ \\
Distance distribution ($\mathcal{U}[d_{\rm min},d_{\rm max}]$) & $\mathcal{U}[10,100]$ m \\
UAV-BS altitude ($h_\mathsf{uav}$) & 100 m \\
Carrier frequency ($f_{\rm c}$) & 28 GHz \\
Bandwidth ($B$) & 100 MHz \\
Noise figure & 9 dB \\
Thermal noise & {-}174 dBm/Hz\\ 
Number of multipaths ($L_\mathsf{p}$) & 5 \\
Angular spread ($\Delta$) & $10^{\circ}$ \\
AoD/AoA distribution & Laplace \\\hline
\hline
\end{tabular}
\vspace{-0.15in}
\end{table}

In this section, we propose a non-linear precoding scheme for the data and AN signal to further improve secrecy rates. Towards this end, we first consider the optimization problem that aims at maximizing sum secrecy rate, which is given as
\begin{IEEEeqnarray}{rl}
    \max_{\textbf{W},\textbf{V}}
    &\quad \sum_{k=1}^{K}\left[\log\left(\frac{1+\mathsf{SINR}_k}{1+\mathsf{SINR}_\mathsf{e}}\right)\right]^+,  \label{eq:optimization_ssr_1}\\
    \text{s.t.}
    &\quad \sum^K_{k=1}\|\textbf{w}_k\|^2+\sum^Z_{i=1}\|\textbf{v}_i\|^2 \leq 1, \IEEEyessubnumber \label{eq:optimization_ssr_2}
\end{IEEEeqnarray}
which is obviously non-convex, and requires sophisticated numerical optimization methods to solve. We therefore propose an alternative optimization problem which aims at minimizing the transmitted power while satisfying particular SINR constraints for each user and Eve. The respective optimization problem is based on the \textit{SINR balancing} strategy \cite{Ottersten2010ConOpt}, and the proposed optimization scheme is described as follows
\begin{IEEEeqnarray}{rl}
    \min_{\textbf{W},\textbf{V}}
    &\quad \sum^K_{k=1} \left\|\textbf{w}_k \right\|^2 + \sum^Z_{i=1} \left\| \textbf{v}_i \right\|^2 , \label{eq:optimization_1}\\
    \text{s.t.}
    &\quad \mathsf{SINR}_k \geq \gamma_k, \, \forall k \in \mathcal{K}, \IEEEyessubnumber \label{eq:optimization_2}\\
    &\quad \mathsf{SINR}_\mathsf{e}\leq\gamma_\mathsf{e}, \IEEEyessubnumber \label{eq:optimization_3}
\end{IEEEeqnarray}
where $\gamma_k$ and $\gamma_\mathsf{e}$ represent the target SINR for the $k$-th user and Eve, respectively, to achieve a specific secrecy rate, and, hence, can be interpreted as \textit{quality-of-service (QoS)} metrics for the underlying secure transmission scheme. Note that although the original SINR balancing problem of \cite{Ottersten2010ConOpt} considers only the minimum SINR threshold for the legitimate users (i.e., $\gamma_k$'s in \eqref{eq:optimization_2}), our strategy employs also the \textit{maximum} SINR threshold for Eve (i.e., $\gamma_\mathsf{e}$ in \eqref{eq:optimization_3}). In addition, our objective function in \eqref{eq:optimization_1} involves the precoders for not only the data but also the AN signal, which makes the solution even more difficult.   

In an attempt to obtain a solution to the optimization problem in \eqref{eq:optimization_1}, we assume that the AN precoder is available a priori. The desired optimization can then be written for the unnormalized data precoder matrix $\Tilde{\textbf{W}}$ with the help of \eqref{eq:SINR_u} and \eqref{eq:SINR_eve} as follows
\begin{IEEEeqnarray}{rl}\label{eq:opti}
    \min_{\Tilde{\textbf{W}}}
    &\quad \sum^K_{k=1} \left\| \Tilde{\textbf{w}}_k \right\|^2, \label{eq:optimization_4} \\
    \text{s.t.}
    &\quad \left| \textbf{h}_k^{\rm H}\Tilde{\textbf{w}}_k \right|^2 \geq \gamma_k \Big( \left\|\textbf{h}_k^{\rm H} \Tilde{\textbf{W}}_{\overline{k}} \right\|^2_\mathsf{F} + \left\|\textbf{h}_k^{\rm H} \textbf{V} \right\|^2 + \sigma^2_n \Big),  \IEEEeqnarraynumspace \IEEEyessubnumber \label{eq:optimization_5}\\
    &\quad \left\| \textbf{H}_e^{\rm H}\Tilde{\textbf{w}}_k \right\|^2 \leq \gamma_\mathsf{e} \left(  \left\|\textbf{H}_e^{\rm H}\textbf{V} \right\|^2_\mathsf{F} + \sigma^2_n \right), \forall k \in \mathcal{K}, \IEEEeqnarraynumspace \IEEEyessubnumber \label{eq:optimization_6}
\end{IEEEeqnarray}
where $\Tilde{\textbf{W}}_{\overline{k}}$ is obtained by removing the $k$-th column from $\Tilde{\textbf{W}}$. The objective function in \eqref{eq:optimization_4} is a quadratic function of the data precoders, and is therefore convex. In addition, \eqref{eq:optimization_6} is also a convex function of the data precoders for a given AN precoder matrix $\textbf{V}$. 

Note that any phase rotation of the optimal data precoder $\textbf{w}_k^\mathsf{opt}$, which is represented by $\textbf{w}_k^\mathsf{opt} e^{j\varphi_k}$ for arbitrary $\varphi_k \,{\in}\, [0,2\pi]$, still achieves the optimal solution since such phase phase ambiguities are eliminated by the absolute value operators in \eqref{eq:SINR_u}. In order to reveal the convexity of the constraint in \eqref{eq:optimization_5}, we make use of this observation by choosing the arbitrary $\varphi_k$ such that the effective channel gain of the $k$-th user, denoted by $\textbf{h}_k^{\rm H}\Tilde{\textbf{w}}_k$, becomes \textit{real-valued} and \textit{nonnegative}, i.e., $\textbf{h}_k^{\rm H}\Tilde{\textbf{w}}_k \,{\geq}\, 0$ and $\operatorname{Im} \{ \textbf{h}_k^{\rm H}\Tilde{\textbf{w}}_k \} \,{=}\, 0$. The respective optimization problem is accordingly expressed as 
\begin{IEEEeqnarray}{rl}
    \min_{\Tilde{\textbf{W}}}
    &\quad \sum^K_{k=1} \left\| \Tilde{\textbf{w}}_k \right\|^2, \label{eq:optimization_7}\\
    \text{s.t.}
    &\quad \textbf{h}_k^{\rm H}\Tilde{\textbf{w}}_k \geq \sqrt{\gamma_k} \left\| \textbf{h}_k^{\rm H}\Tilde{\textbf{W}}_{\overline{k}} \;\;\; \textbf{h}_k^{\rm H}\textbf{V} \;\;\; \sigma_n \right\|, \IEEEeqnarraynumspace \IEEEyessubnumber \label{eq:optimization_8}\\
    &\quad \left\| \textbf{H}_e^{\rm H}\Tilde{\textbf{w}}_k \right\| \leq \sqrt{\gamma_\mathsf{e} \left( \left\|\textbf{H}_e^{\rm H} \textbf{V} \right\|^2_\mathsf{F} + \sigma^2_n \right)} , \IEEEeqnarraynumspace \IEEEyessubnumber \label{eq:optimization_9}\\
    &\quad \textbf{h}_k^{\rm H}\Tilde{\textbf{w}}_k \geq 0 , \; \operatorname{Im} \left( \textbf{h}_k^{\rm H} \Tilde{\textbf{w}}_k \right) = 0 , \; \forall k \in \mathcal{K}, \IEEEeqnarraynumspace \IEEEyessubnumber \label{eq:optimization_10}
\end{IEEEeqnarray}
which is in the form of second-order cone programming (SOCP), and can be solved by standard convex-optimization solvers (e.g., MATLAB CVX). The solution of \eqref{eq:optimization_7} is still cumbersome in terms of picking up appropriate SINR thresholds $\gamma_k$ and $\gamma_\mathsf{e}$ for $k \,{\in}\, \mathcal{K}$. As a practical yet efficient solution, we obtain these thresholds from the $\mathsf{SINR}_k$ and $\mathsf{SINR}_\mathsf{e}$ results associated with the RZF precoding given in \eqref{eq:proposed_rzf}. Note that the final data precoders with normalized power are calculated by \eqref{eq:proposed_data_precoder_normalized}, and the initial AN precoder is obtained by \eqref{eq:proposed_eve_precoder_normalized}.

\begin{figure}[!t]
	\centering
	\subfloat[$K=4$ (Underloaded)]{
	\includegraphics[width=0.435\textwidth]{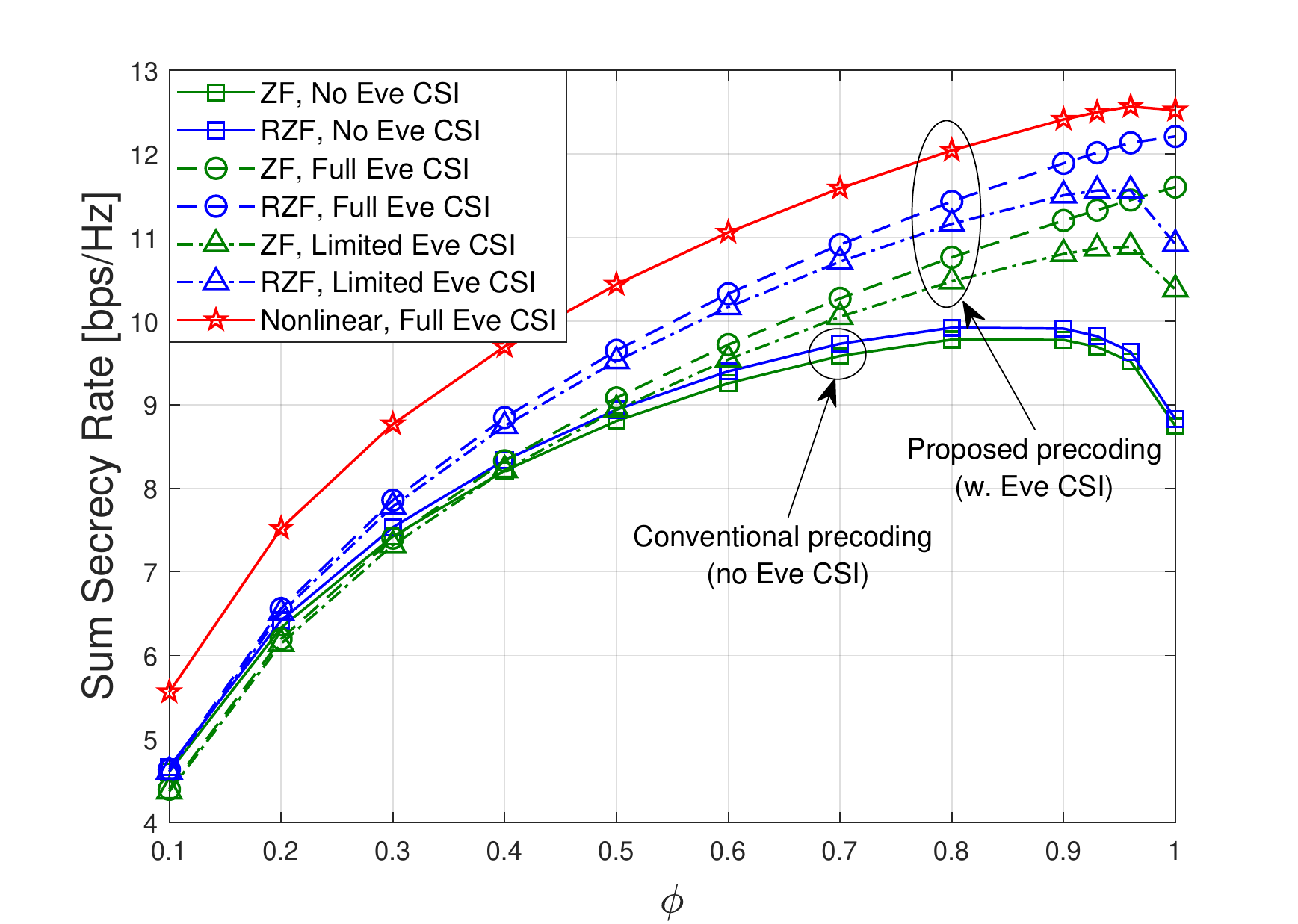}\label{fig:srate_K4}}
	\vspace{-0.0in}
	\subfloat[$K=16$]{
	\includegraphics[width=0.435\textwidth]{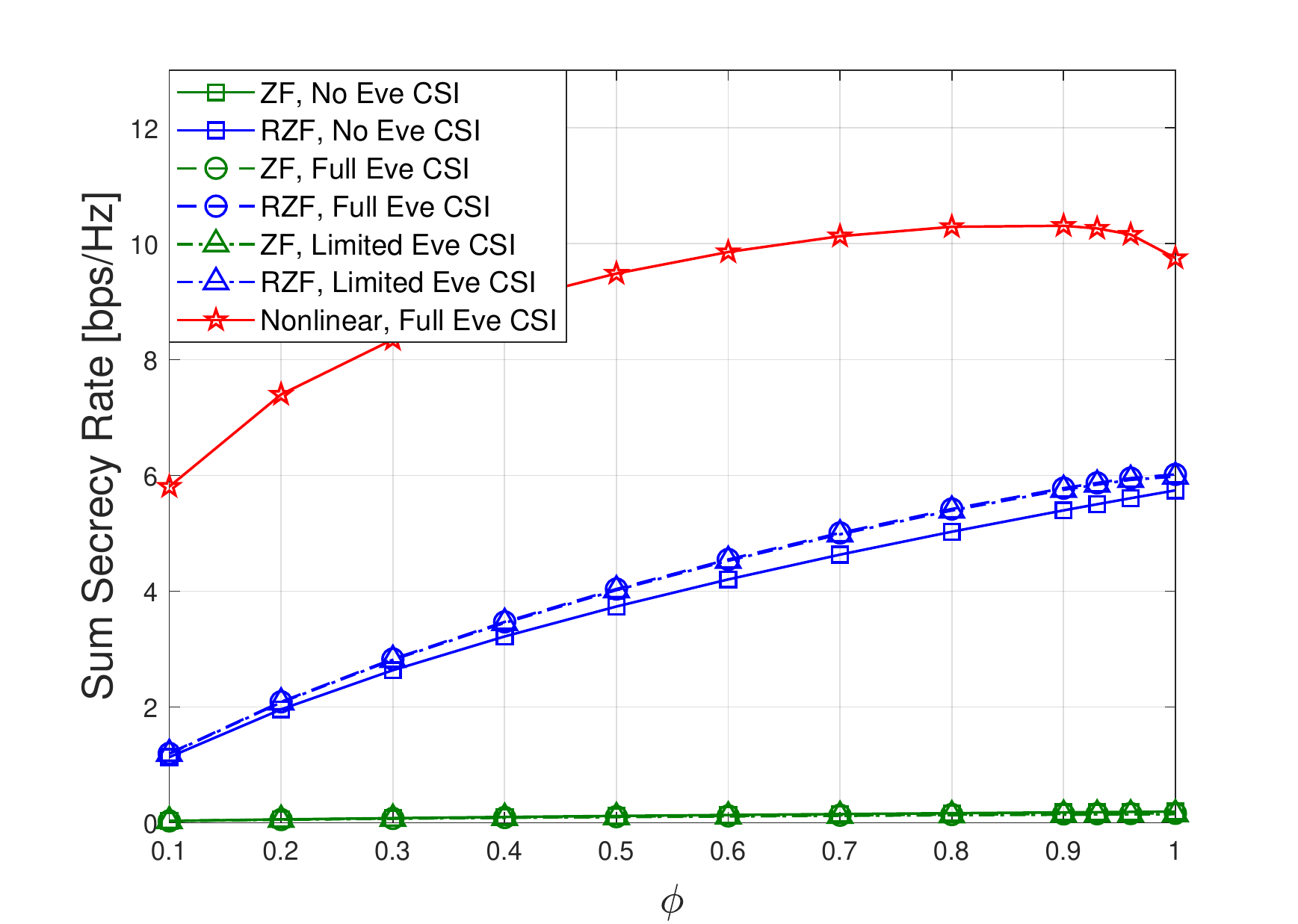}\label{fig:srate_K16}}
	\vspace{-0.0in}
	\subfloat[$K=32$ (Overloaded)]{
	\includegraphics[width=0.435\textwidth]{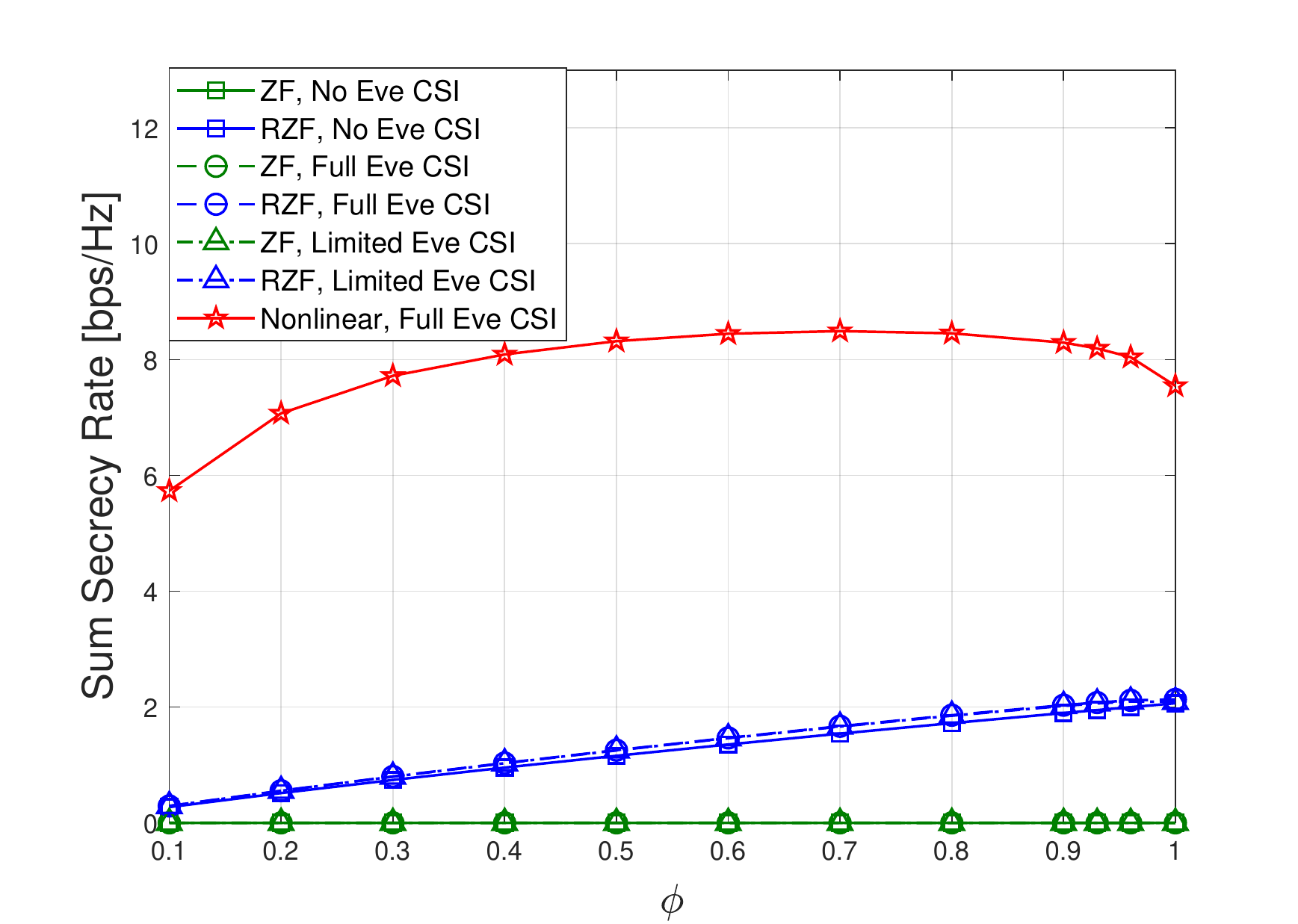}\label{fig:srate_K32}}
	\vspace{-0.0in}
	\caption{Sum secrecy rate vs power splitting factor $\phi$ for $\mathsf{P}_\mathsf{Tx} \,{=}\, 30\,\text{dBm}$, $N\,{=}\,16$, $M\,{=}\,4$, and $K \,{\in}\, \{4,16,32\}$.} 
	\label{fig:srate_K}
	\vspace{-0.3in}
\end{figure}

\section{Numerical Results} \label{sec:results}

In this section, we present the numerical results based on extensive Monte Carlo simulations to evaluate the performance of the proposed precoding strategies in comparison to the conventional designs. We adopt the urban micro (UMi) mmWave path-loss model by 3GPP given (in dB scale) as follows \cite{3gpp} 
\begin{align}\label{eq:PL}
    \mathsf{PL} \left( d_k \right) &= 32.4 + 21\log_{10}\left( d_k^\mathsf{\,LoS} \right) + 20\log_{10}\left( f_{\rm c} \right),
\end{align}
where $d_k^\mathsf{\,LoS} \,{=}\, \sqrt{d_k^2+h_\mathsf{uav}^2}$ is the LoS distance of the $k$-th user, and $f_{\rm c}$ is the carrier frequency (normalized by $1$ GHz). The complete list of the simulation parameters are given in Table~\ref{table:settings}.

In Fig.~\ref{fig:srate_K}, we depict sum secrecy rates along with varying power splitting factor such that $\phi \,{\in}\, [0,1]$, $N \,{=}\, 16$, $M \,{=}\, 4$, and $K \,{\in}\, \{4, 16, 32\}$. We observe that the proposed nonlinear precoder design is superior to both the conventional and proposed linear precoding strategies. In addition, the proposed linear precoding schemes involving either ZF or RZF (\textit{with} Eve CSI) achieve a better sum-rate performance as compared to the conventional linear schemes (\textit{without} Eve CSI). Note that this performance improvement in linear precoders (associated with the use of Eve CSI) diminishes as the number of users increases, and becomes marginal in the overloaded scenarios (e.g.,  Fig.~\ref{fig:srate_K}\subref{fig:srate_K32}). Note also that, the performance of ZF precoder with or without Eve CSI deteriorates rapidly along with increasing number of users, which directly follows from the multiplication $\textbf{H}^{\rm H} \textbf{H}$ in \eqref{eq:precoder_zf_unnormalized} being not invertible.

The sum-rate performance in Fig.~\ref{fig:srate_K} is observed to follow generally an increasing trend along with the power splitting factor $\phi$. In particular, the conventional linear precoders (without Eve CSI) have the best secrecy-rate performance around $\phi \,{\in}\,[0.8,0.9]$ for the \textit{underloaded} scenario of $K \,{=}\, 4$, as shown in Fig.~\ref{fig:srate_K}\subref{fig:srate_K4}. On the other hand, the proposed linear and nonlinear precoders with \textit{full} Eve CSI achieves the best sum secrecy rate for $\phi \,{=}\, 1$ under the same setting. The proposed linear precoders with \textit{limited} Eve CSI, however, have their peak corresponding secrecy-rate performance at $\phi \,{\in}\,[0.9,1.0]$. These observations conclude that incorporating the Eve CSI in the precoder design (as in the proposed linear and nonlinear precoders) relieve the necessity of the AN transmission for underloaded scenarios. In particular, the use of full Eve CSI completely eliminates any need for AN (i.e., $\phi^\mathsf{opt} \,{=}\, 1$) while the limited Eve CSI still requires small amount of AN transmission (i.e., $\phi^\mathsf{opt} \,{<}\, 1$). Note also that limited Eve CSI has a very close performance to full Eve CSI for proposed linear precoders in underloaded scenarios (except still requiring AN transmission). 

As the number of users increases and the setting becomes overloaded (e.g., Fig.~\ref{fig:srate_K}\subref{fig:srate_K32}), the linear precoders cannot provide a promising secrecy-rate performance irrespective of using any Eve CSI or not at all. The proposed nonlinear precoder, however, has a significantly better sum secrecy rates with a peak at $\phi \,{\in}\,[0.6,0.8]$. As a result, although the proposed nonlinear precoder achieves a promising sum-secrecy rate for overloaded scenarios, it is not possible to get rid of AN transmission completely (in contrast to underloaded scenarios). 

\section{Conclusion}\label{sec:conclusion}

We study physical layer security for UAV-BS communications in mmWave frequency spectrum which considers transmission of AN to improve secrecy rates. We propose linear and nonlinear precoding schemes which takes into account full or limited feedback on the Eve CSI. The proposed nonlinear precoding scheme is shown to outperform any other candidate schemes, which is capable of eliminating any necessity for AN transmission in underloaded scenarios. The proposed linear precoding scheme is also shown to be very effective as compared to the conventional schemes, especially for underloaded scenarios. 
\vspace{-0.1in}

\bibliographystyle{IEEEtran} 
\bibliography{IEEEabrv,bibfile}

\end{document}